\documentclass[prb,aps,twocolumn]{revtex4}
\usepackage{graphicx}
\usepackage{amsmath}

\begin{document}

\title{Superconducting Fluctuation Corrections to the Thermal Conductivity
in Granular Metals}

\author{C. Biagini}
\affiliation{INFM, Unit\`a di Firenze, Via G. Sansone 1,
    50019 Sesto F.no, Firenze, Italy}
\author{R. Ferone}
\affiliation{LPMMC, CNRS \& Universit\'e Joseph Fourier,
    BP 166, 38042 Grenoble CEDEX 9, France}
\affiliation{NEST-INFM \& Scuola Normale Superiore,
    I-56126 Pisa, Italy}
\author{Rosario Fazio}
\affiliation{NEST-INFM \& Scuola Normale Superiore,
    I-56126 Pisa, Italy}
\author{F.W.J. Hekking}
\affiliation{LPMMC, CNRS \& Universit\'e Joseph Fourier,
    BP 166, 38042 Grenoble CEDEX 9, France}
\author{V. Tognetti}
\affiliation{INFM, Unit\`a di Firenze, Via G. Sansone 1,
    50019 Sesto F.no, Firenze, Italy}
\affiliation{Dipartimento di Fisica,
    Universit\`a di Firenze, Via G. Sansone 1, 50019 Sesto F.no,
    Firenze, Italy}
\affiliation{INFN, Sezione di Firenze, Via
    G. Sansone 1, 50019 Sesto F.no, Firenze, Italy}

\date{\today}

\begin{abstract}
The first-order superconducting fluctuation corrections to the thermal conductivity of a
granular metal are calculated. A suppression of thermal conductivity
proportional to $T_c/(T-T_c)$ is observed in a region not too
close to the critical temperature $T_c$. As $T\simeq T_c$, a saturation
of the correction is
found, and its sign depends on the ratio between the
barrier transparency and the critical temperature. In both regimes, the
Wiedemann-Franz law is violated.
\end{abstract}

\maketitle

\section{Introduction}

In normal metals, in the presence of BCS  interaction, electrons
can form Cooper pairs even for temperatures $T$ larger than the
critical temperature $T_c$. As $T \ge T_c$, the pairs have a
finite lifetime, the Ginzburg-Landau (GL) time, inversely
proportional to the distance from the critical temperature
$\tau_{GL} \sim \left(T-T_c\right)^{-1}$. These superconducting
fluctuations strongly affect both the thermodynamic and transport
properties and since many years they are widely studied both
theoretically and experimentally~\cite{LV04}.\\ \indent The first
analysis of fluctuation corrections has been performed on
electrical conductivity where the pairing leads to three distinct
contributions named the Aslamazov-Larkin (AL), the Maki-Thompson
(MT) and Density of States (DOS) terms. In the first one, the
formation of Cooper pair leads to a parallel superconducting
channel in the normal phase; the second takes into account the
coherent scattering off impurities of the (interacting) electrons;
finally, the third one is due to the rearrangement of the states
close to the Fermi energy since electrons involved in pair
transport are no longer available for single particle transport.
Both the AL and MT terms lead to an enhancement of the
conductivity above $T_c$, on the contrary, the DOS correction is
of opposite sign.\\ \indent 
The analysis of superconducting
fluctuation corrections to thermal conductivity dates back to the
early 1960s, when Schmid~\cite{S66} and Caroli and
Maki~\cite{CM67} found an expression for the heat current in the
framework of the phenomenological time dependent GL theory,
(TDGL). More recently, a complete analysis was performed, 
in the same framework of the TDGL, by Ussishkin\cite{ISS03}. 
Abrahams {\em et al.}~\cite{ARW70}  first pointed out the divergence
of the thermal conductivity in the vicinity of the critical
temperature due to the opening of the fluctuation pseudogap in the
density of states (DOS) energy dependence in the homogeneous case.
Niven and Smith have shown~\cite{NS02} that Abrahams's DOS
correction [$\approx Gi\ln(1/\epsilon)$,
$\epsilon=(T-T_c)/T_c$, $Gi$ being the so-called
Ginzburg-Levanyuk parameter] is exactly compensated by the regular
Maki-Thompson (MT) one; hence, all singular first order
fluctuation corrections are cancelled out. The only surviving
contribution to heat conductivity, the Aslamazov-Larkin (AL) one,
is non-singular in temperature. Therefore, in bulk metals, no
singular behaviour of the heat current is expected at the
metal-superconductor phase transition.\\ \indent In this paper we
are interested in the superconducting fluctuation corrections to
the thermal conductivity in a granular superconductors, an
ensemble of metallic grains embedded in an insulating amorphous
matrix and undergoing a  metal-superconductor phase transition due
to the existence of pairing interaction inside each grain. The
electrons can diffuse in the system due to tunneling between the
grains. Experimentally, this kind of systems have been
investigated, for example, in Ref. \onlinecite{GER97}. Each Al
grain has an average dimension of 120\AA, while the sample has a
linear dimension of the order of mm, that is, much larger than the
superconducting coherence length. The reason for studying thermal
transport in granular metals is that, depending on the temperature regime, 
a radically
different behaviour, as compared with the homogeneous case, may emerge. In
fact, in granular material (a similar situation occurs in layered
superconductors) the AL and MT contributions are of higher order
in the tunneling amplitude as compared to the DOS. This effect has
been observed, for example, in the electrical~\cite{BMV93} and the
optical conductivity~\cite{FV96} of layered superconductors and
the electrical conductivity~\cite{BEL00} of granular systems.
Indeed, in granular superconductors there is a temperature region
in which  a singular correction due to superconducting
fluctuations for a quasi-zero-dimensional system dominates the
behaviour of the thermal conductivity; such a correction can be
either negative or positive, depending on the ratio between the
barrier transparency and the critical temperature
$T_c$. When the temperature approaches $T_c$, the behaviour
observed in homogeneous systems is recovered, and the divergence
will be cut off to crossover to the regular behaviour. Moreover,
a significant difference with the homogeneous systems is present, 
the constant correction at $T=T_c$ being either negative or
positive depending on the above-mentioned ratio. For some choice of 
the parameter, a
non-monotonic temperature dependent behaviour of the correction is possible.\\
\indent A phenomenological approach to granular superconductors
has been proposed long ago~\cite{P67,DIG74}, while the microscopic
theory has only been formulated very recently~\cite{BEL00,LV04}.
The difference between bulk and granular microscopic theory is
mainly based on the renormalization of the superconducting
fluctuation propagator due to the presence of tunneling. This
renormalization accounts for the possibility that each electron
forming the fluctuating Cooper pair tunnels between neighbor
grains during the Ginzburg-Landau time.\\ \indent 
The paper is
organized as follows. In Section~\ref{model}, we describe and
formulate the model. Section~\ref{fluct} contains the main steps
and assumptions of the calculation of fluctuation propagator. Its
expression, calculated in Ref. \onlinecite{BEL00}, is given
explicitly at every order in tunneling in the ladder
approximation. By means of that, DOS, MT, and AL corrections are
evaluated. For each of those corrections, an explicit form for the
response function is presented. In the final section, we discuss
the overall behaviour of the fluctuation corrections to thermal
conductivity as a function of temperature. For temperatures
sufficiently far from $T_c$, the system behaves as in the zero-dimensional case. 
In this region, the correction to
the heat conductivity  has a singular behaviour:
$\left|\delta\kappa\right|\propto \kappa_0/\left({g_T}\epsilon\right)\;$,
where $\kappa_0$ is the classical Drude conductivity for a
granular metal, and it reads
\begin{equation}
\kappa_0=\frac{8\pi}{3}g_Ta^{2-d}T\;, \label{kappa}
\end{equation}
$a$ being the size of a single grain, $d$ the dimensionality
of the system  and $\epsilon=\left(T-T_c\right)/T_c$ the
reduced temperature. We defined the dimensionless macroscopic
tunneling conductance $g_T=\left[(\pi/2)t\nu_F\right]^2$, with
$\nu_F$ the electronic density of states at the Fermi level, and
$t$ the hopping energy. On the other hand, when the correlation
length increases until the distance between two nearest neighbor
grains, the tunneling becomes important and the correction,
exactly at the critical temperature,  reduces to a constant
\begin{equation}
\delta\kappa=\frac1{zg_T}\left(\frac9{2\pi}\frac{g_T\delta}{T_c}
-\frac3{\pi^2}\right){\kappa_0}. 
\end{equation}
Connections with the
homogeneous metal results are discussed. In the appendix, we
briefly review the evaluation of the superconducting fluctuation
propagator in a granular metal. Throughout the paper, we set
$\hbar=k_B=1$.

\section{The model}\label{model}

We consider a $d-$dimensional array of metallic grains embedded in
an insulating amorphous matrix, with impurities on the surface and
inside each grain. Even if the analytical model we use is for a
perfectly ordered $d$-dimensional matrix, the results we found
still hold for an amorphous one. Indeed, one can imagine different
possible configurations of spatial position of grains in the
lattice, that is, different disordered configurations.
Consequentely, the hopping matrix shall vary for each sample. By
performing the average over disorder, one gets a model with the
same value of coordination number and hopping energy, $t$, for
different configurations. In other words, our description is
correct until the system can be described by a dimensionless
tunneling conductance, $g_T$, on a scale which is much bigger than
the typical linear dimension of the grains, $a$, but smaller than
the macroscopic dimension of the whole sample.
\\
\indent The Hamiltonian of the system reads
\begin{equation}
\label{1} \hat H=\hat H_0+\hat H_P+\hat H_T \;.
\end{equation}
$\hat H_0$ and $\hat H_P$ describe the free electron gas and the
pairing Hamiltonian inside each grain, respectively
\begin{eqnarray}
\label{2} \hat H_0&=&\sum_{i,k}\varepsilon_{i,k}\hat
a^\dagger_{i,k}\hat a_{i,k}+\hat H_\mathrm{imp}\;,
\\
\label{3} \hat H_P&=&-\lambda \sum_{i,kk'} \hat
a^\dagger_{i,k}\hat a^\dagger_{i,-k}\hat a_{i,-k'}\hat a_{i,k'}\;,
\end{eqnarray}
where $i$ is the grain index, and $\hat a^\dagger_{i,k}$ ($\hat
a_{i,k}$) stands for creation (annihilation) operator of an
electron in the state $k=\left({\bf k},\uparrow\right)$ or
$-k=\left(-{\bf k},\downarrow\right)$. The term $\hat H_\mathrm{imp}$
describes the electron elastic scattering with impurities. The
interaction term in Eq.(\ref{1}) contains only diagonal
terms\cite{ABP00}. Such a description is correct in the limit
\begin{equation}
\delta\ll\Delta\ll E_T \;,\label{dis}
\end{equation}
where $\delta\sim\nu_F^{-1}$ is the mean level spacing and the
smallest energy scale in the problem, and $\Delta$ the (BCS)
superconducting gap of a single grain, supposed equal for each of
them. $E_T=D/{a^2}$ is the Thouless energy, $D$ being the
intragrain diffusion constant. Under the previous assumption,
Eq.(\ref{dis}), one can safely neglect off-diagonal $1/g$
corrections, where $g$ is the dimensionless conductance of a
grain, $g={E_T}/\delta$. Equation (\ref{dis}) is equivalent to the
condition $a\ll \xi_0$, where $\xi_0=\sqrt{D/T_c}$ is  the dirty
superconducting coherence length; then, Eq.(\ref{1}) describes an
ensemble of zero-dimensional grains. In addition, Eq.(\ref{dis})
states that the energy scale, $\tau^{-1}$, with $\tau$ being the mean
free time, related to $\hat H_\mathrm{imp}$ is much larger than $\Delta$.\\
\indent The grains are coupled by tunneling. The tunneling
Hamiltonian is written as ($t\ll E_T$)
\begin{equation}
\label{4} \hat H_T= \sum_{\left\langle i,j\right\rangle}\sum_{{\bf
p}{\bf q},\sigma}\left[t_{ij}^{pq}\hat a^\dagger_{i,{\bf
p}\sigma}\hat a_{j,{\bf q}\sigma}+\mathrm{H.c}\right]\;.
\end{equation}

\begin{figure}[t]
\begin{center}
\includegraphics[width=6cm]{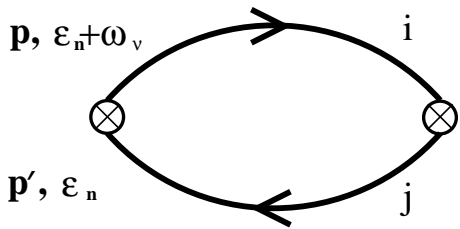}
\end{center}
\caption{Diagram for the thermal conductivity in granular metals.
The solid lines are impurity-averaged single-electron Green's functions with the specified
momentum and Matsubara's frequency, and belonging to the grain $i$
and $j$. The vertices are discussed in the text.} \label{fig1}
\end{figure}

We assume that the momentum of an electron is completely
randomized after the tunneling. Finally, assuming that the system
is macroscopically a good metal, $t\gg\delta$, we can safely
neglect the Coulomb interaction, it being well
screened\cite{nota3}, and weak localization corrections too,
at least for not too low temperatures\cite{BCTV05}, i.e. when $T\lesssim g_T\delta$.\\ \indent
The tunneling heat current operator is given as
\begin{equation}
\label{5} \hat j^\mathrm{(heat)}= ia\sum_{\left\langle
i,j\right\rangle}\sum_{{\bf p}{\bf q}\sigma}\left[\varepsilon_n
t_{ij}^{pq} \hat a_{i,{\bf p}\sigma}^\dagger \hat a_{j,{\bf
q}\sigma}-\mathrm{H.c.}\right]\;,
\end{equation}
where $\varepsilon_n$ is the Matsubara frequency of the electron involved
in the transport.\\
\indent In linear response theory, the heat
conductivity is defined as
\begin{equation}
\label{7} \kappa=\lim_{\omega\rightarrow0}\left[\frac{{\cal
Q}^\mathrm{(heat)}_\mathrm{ret}\left(i\omega_{\nu}\right)}{\omega_{\nu}
T}\right]_{i\omega_{\nu}\rightarrow\omega+i0^+}\;,
\end{equation}
where ${\cal Q}^\mathrm{(heat)}(\omega_{\nu})$ is the linear response
function to an applied temperature gradient:
\begin{eqnarray}
\lefteqn{{\cal
Q}^\mathrm{(heat)}(\omega_{\nu})=Tt^2a^2\sum_{\left\langle
i,j\right\rangle}\sum_{\varepsilon_n}\left(\varepsilon_{n+\nu}+\varepsilon_{n}
\right)^2}\nonumber\\
& &\times\int(dp)G\left(\tilde\varepsilon_{n+\nu},{\bf
p}\right)\int(dq)G\left(\tilde\varepsilon_{n},{\bf q}\right)\;,
\label{6}
\end{eqnarray}
where $G\left(\tilde\varepsilon_{n},{\bf p}\right)$ is the exact
Matsubara Green's function of an electron in a grain,
$(dp)=[d^dp/(2\pi)^d]$, $\tilde\varepsilon_n$ and
$\varepsilon_{n+\nu}$ are shorthand notations for
$\varepsilon_n+(i/{2\tau}){\rm sign}\varepsilon_n$ and
$\varepsilon_n+\omega_{\nu}$, respectively. In the latter
equation, we considered the tunneling amplitude uniform and
momentum independent, $t_{ij}^{pq}\equiv t$.
The thermal conductivity for free electrons, $\kappa_0$, Eq.
(\ref{kappa}), is given by the diagram in Fig. \ref{fig1}, where,
as usual, Green's function is $G\left(\tilde\varepsilon_{n},{\bf
p}\right)=1/[i\tilde\varepsilon_{n}-\xi({\bf p})]$, and each
vertex contributes as $i2at(\varepsilon_n+\omega_\nu/2)$.
Electrical conductivity reads $\sigma_0=e^2(8/\pi) g_Ta^{2-d}$;
therefore, the Lorenz number is
$L_0={\kappa_0}/{\sigma_0T}={\pi^2}/{3e^2}$.
\section{Superconducting fluctuation corrections to thermal
conductivity}
\label{fluct}
At temperatures above but not far from the critical one,
superconducting fluctuations allow the creation of Cooper pairs
that strongly affect transport. In other words, fluctuations open
a new transport channel, the so-called {\em Cooper pair
fluctuation propagator}, Ref. \onlinecite{LV04}. It is such a
contribution that gives rise to corrections to both
the electrical and thermal conductivity.\\
\indent With respect to the bulk case, the propagator is
renormalized by the tunneling, and as explained in the appendix,
it takes into account the possibility that each electron forming
the cooper pair
can tunnel from one grain to another, without loosing the coherence.\\
\indent
The expression for the superconducting fluctuation propagator for
a granular metal, calculated in Ref. \onlinecite{BEL00}, is:
\begin{eqnarray}
\label{18} \Lambda_{\bf K}\left(\Omega_{\mu}\right)=-\frac1{\nu_F}
\frac1{\ln\frac
T{T_c}+\frac{\pi\left|\Omega_{\mu}\right|}{8T_c}+z\frac{g_T\delta}{T_c}
\left(1-\gamma_{\bf K}\right)},
\end{eqnarray}
where ${\bf K}$ is the wave vector associated with the lattice of
the grains, $\Omega_{\mu}$ is a bosonic Matsubara's frequency, and
$z$ the number of nearest neighbor grains. The function
$\gamma_{\bf K}=(1/z)\sum_{\bf a}e^{i{\bf K}\cdot{\bf a}}$ is
the so-called lattice structure factor, where ${\bf a}$ is a
vector connecting nearest neighbor grains. The main steps of the
calculation of Eq. (\ref{18}), done in Ref. \onlinecite{BEL00},
are reviewed for completeness in the appendix.

\begin{figure}[t]
\begin{center}
\includegraphics[width=5cm]{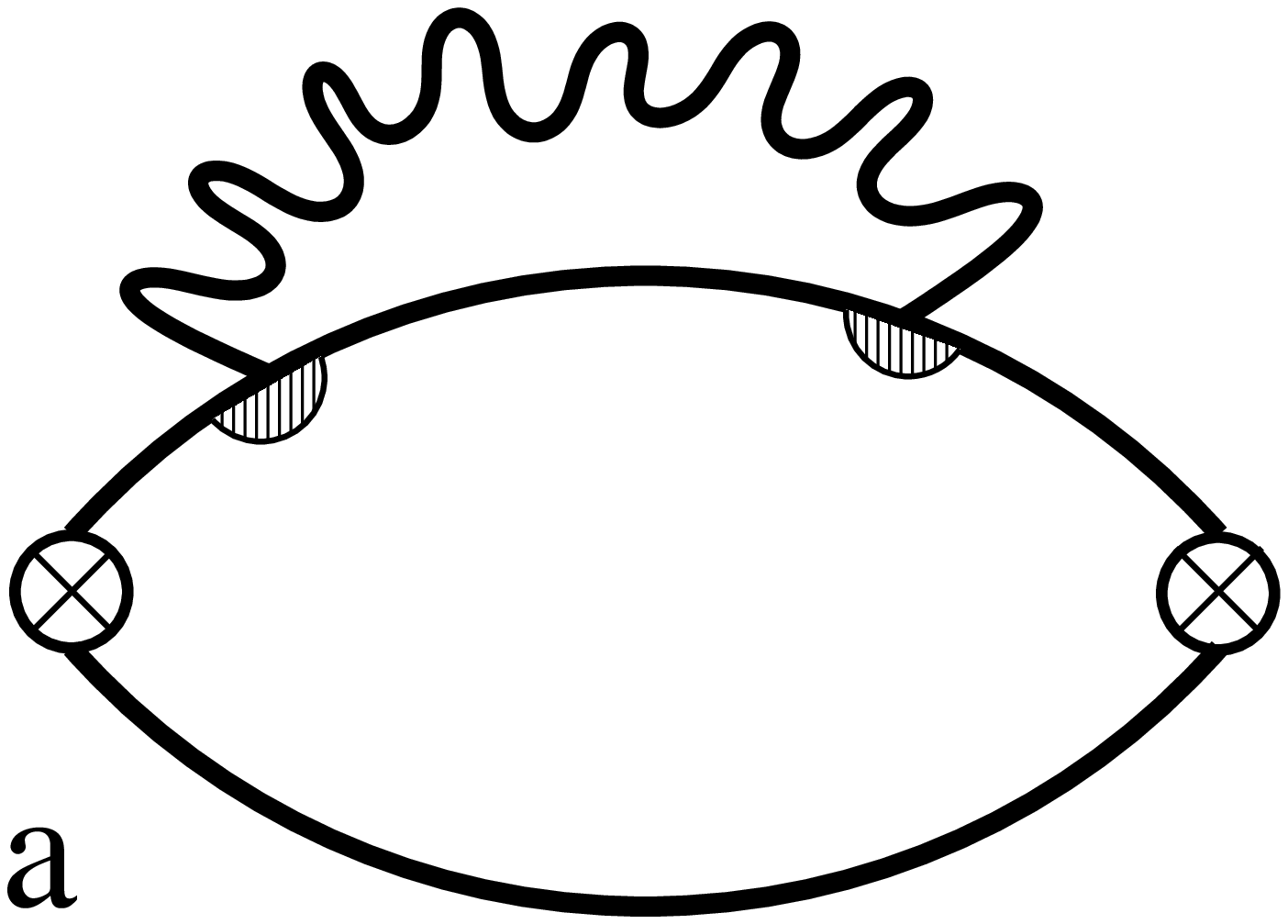}
\includegraphics[width=5cm]{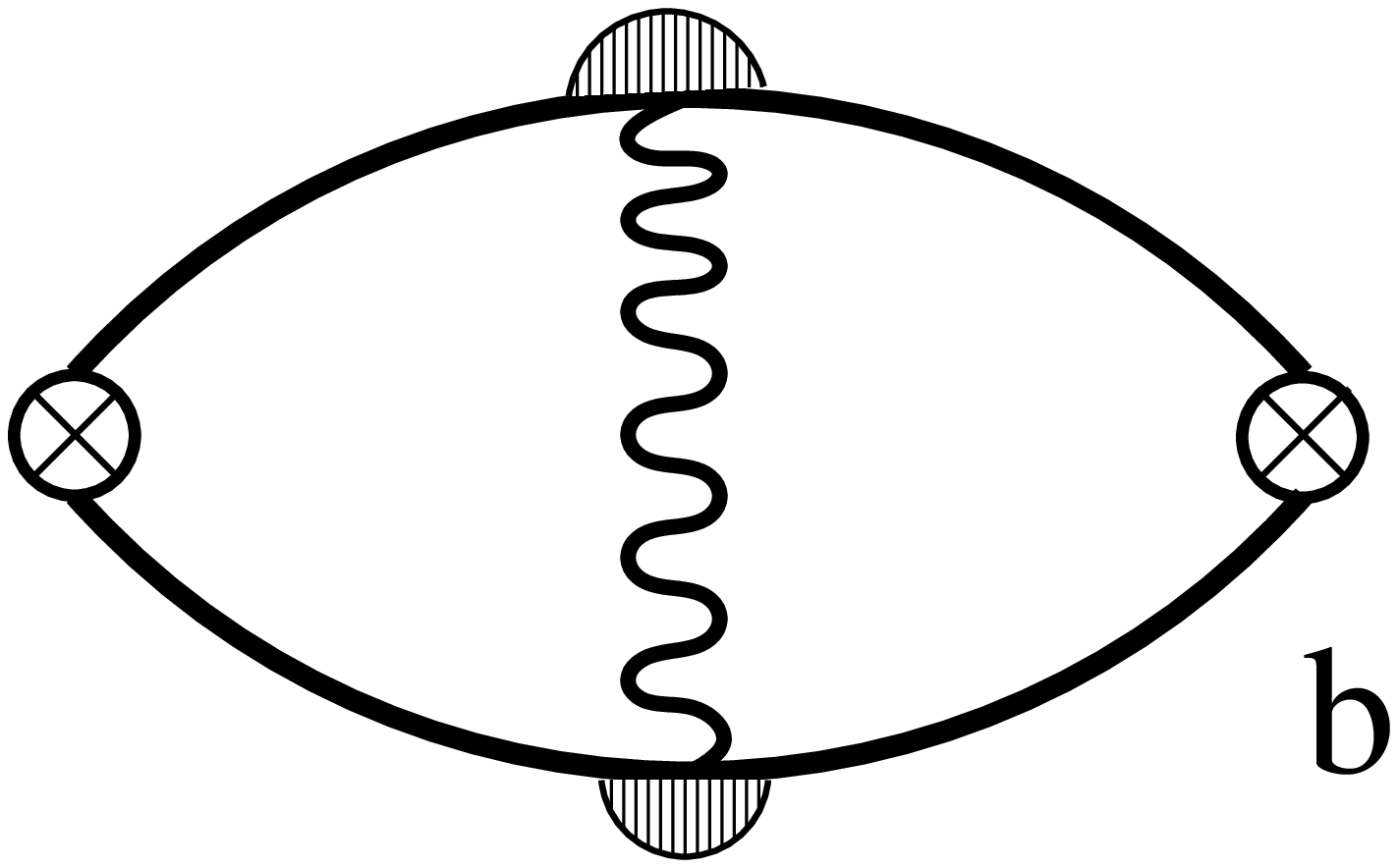}
\end{center}
\caption{Diagrams representing various fluctuation contributions
to the thermal conductivity. (a) density of states contribution.
(b) Maki-Thompson contribution. The solid lines are
impurity-averaged single-electron Green's functions, wavy lines
represent the fluctuation propagator and the shadowed areas are
Cooperon vertex corrections. Crossed circles represent tunneling
vertices.} \label{figtot}
\end{figure}

The various contributions to thermal conductivity are shown in
Figs. \ref{figtot} and \ref{figtot1}.

The correction due to the density of states renormalization, Fig.
\ref{figtot}(a), is the only one which is present even in absence of
tunneling; therefore, for temperatures $T-T_c\gg g_T\delta$, 
we expect this term to give a significant contribution to the thermal conductivity. For lower temperatures,
the bulk behaviour will be recovered.\\
\indent The MT correction, represented in Fig. \ref{figtot}(b), can
be evaluated using the same procedure as in the case of the DOS
one. It is important to stress that the sign of linear response
function is the same as for the DOS: in fact, the energies of
electrons entering the diagram from opposite sides have opposite
signs but the same happens to their velocities. In the case of
electrical conductivity, the sign of linear response function is
opposite. It is this difference that ultimately results in the
cancellation of two identical contributions in the
thermal conductivity\cite{NS02}.\\
\indent Let us finally comment on the AL contribution, given by
the diagrams in Fig. \ref{figtot1}. It is well known, in the case
of homogeneous metals, that such a correction to the thermal
conductivity is not singular~\cite{NS02,USH02}. We will show
briefly that in the case of granular metals this correction
vanishes~\cite{nota2} in the static limit too, but not in the
dynamical one, giving an important and characteristic
contribution to the total correction.\\
\indent In the following paragraphs, we present the evaluation of
corrections to thermal conductivity due to different diagrams.

\subsection{Density of states correction}

The diagram for the DOS correction is given in Fig.
\ref{figtot}(a) and the corresponding response function can be written as
\begin{eqnarray}
\label{19} {\cal
Q}^\mathrm{(DOS)}\left(\omega_{\nu}\right)=T^2t^2a^2\sum_{\left\langle
i,j\right\rangle}\sum_{\Omega_{\mu}}\Lambda_{ij}\left(\Omega_{\mu}\right)
\Sigma\left(\Omega_{\mu},\omega_{\nu}\right)\;,
\end{eqnarray}
where
\begin{eqnarray}
\Sigma\left(\Omega_{\mu},\omega_{\nu}\right)&=&
\sum_{\varepsilon_n}\lambda^2\left(\varepsilon_{n+\nu},\varepsilon_{-n-\nu+\mu}\right)
\left(\varepsilon_n+\varepsilon_{n+\nu}\right)^2\nonumber\\
& &\times I\left(\varepsilon_n,\Omega_{\mu},\omega_{\nu}\right)\;,
\label{20}
\end{eqnarray}
and
\begin{eqnarray}
I\left(\varepsilon_n,\Omega_{\mu},\omega_{\nu}\right)&=&
\int\left(dp\right)G^2_0\left({\bf p},\varepsilon_{n+\nu}\right)
G_0\left({\bf p},\varepsilon_{-n-\nu+\mu}\right)\nonumber \\
&&\times \int\left(dp'\right)G_0\left({\bf
p}',\varepsilon_n\right).\label{21}
\end{eqnarray}
We introduced the Cooperon vertex correction,
$\lambda\left(\varepsilon_1,\varepsilon_2\right)=
1/\tau{\left|\varepsilon_1-\varepsilon_2\right|}$ in the
zero-dimensional limit and without tunneling
corrections\cite{nota1}. The main contribution to singular
behaviour comes from ``classical'' frequencies, $|\Omega_{\mu}|\ll
T_c$: consequently, we will take the so-called static limit,
$\Omega_{\mu}=0$, in the calculation of correction. This will be
true also for the Maki-Thompson correction in the next paragraph.
In the dirty limit, we can neglect all the energy scales in the
electronic Green's function in comparison with $1/\tau\gg T$,
and the factor $I\left(\varepsilon_n,0,\omega_{\nu}\right)$ turns
out to be
\begin{equation*}
I\left(\varepsilon_n,0,\omega_{\nu}\right)=-2\left(\pi\nu_F\tau\right)^2
\left[\theta\left(\varepsilon_n\varepsilon_{n+\nu}\right)-
\theta\left(-\varepsilon_n\varepsilon_{n+\nu}\right)\right]\;.
\end{equation*}
\indent Inserting the previous expression in Eq.(\ref{20}), we are left with the
sum over the electronic Matsubara frequencies. It is
straightforward to check that the only contribution linear in
$\omega_{\nu}$ is given by
$\Sigma\left(0,\omega_{\nu}\right)=-\omega_{\nu}\pi\nu_F^2$. By
means of Eq.(\ref{19}), we obtain the general form for
the DOS response function after the analytical continuation
\begin{eqnarray}
\label{22} {\cal
Q}^\mathrm{(DoS)}\left(-i\omega\right)=\left(-i\omega\right)\frac8\pi
g_TTa^2\sum_{\left\langle
i,j\right\rangle}\Lambda_{ij}\left(0\right)\;,
\end{eqnarray}
where we also took into account the multiplicity of the DOS
diagrams. The corresponding correction to heat conductivity is
given by
\begin{eqnarray}
\label{23}
\frac{\delta\kappa^{(DoS)}}{\kappa_0}=-\frac3{\pi^2}\frac{1}{g_T}\frac{g_T\delta}{T_c}
\int_{BZ}(dK)\frac1{\epsilon+z\frac{g_T\delta}{T_c}\left(1-\gamma_{\bf
K}\right)}.
\end{eqnarray}
\indent We took the lattice Fourier transform and defined the
reduced temperature $\epsilon=\ln(
T/{T_c})\simeq{(T-T_c)}/{T_c}$. $(dK)=[{a^d}/{(2\pi)^d}]d^dK$ is
the dimensionless measure of the first Brillouin zone. Close to
$T_c$, the integral takes its main contribution from the small
momentum region and we recover the bulk DOS behaviour as
\begin{eqnarray}
\label{25} \delta\nu\propto\frac{1}{g_T}\left\{
\begin{array}{cc}
\sqrt{\epsilon}, & d=3; \\
\ln\frac1\epsilon, & d=2; \\
\frac1{\sqrt\epsilon}, & d=1.
\end{array}
\right.
\end{eqnarray}

\subsection{Maki-Thompson correction}

The MT correction, [Fig. \ref{figtot}(b)], reads
\begin{eqnarray}
\label{MT1} {\cal
Q}^\mathrm{(MT)}\left(\omega_{\nu}\right)=a^2Tt^2\sum_{\left\langle
i,j\right\rangle}\sum_{\Omega_{\mu}}\Lambda_{ij}\left(\Omega_{ij}\right)
\Sigma\left(\Omega_{\mu},\omega_{\nu}\right)\;,
\end{eqnarray}
where
\begin{eqnarray}
\Sigma\left(\Omega_{\mu},\omega_{\nu}\right)&=&
T\sum_{\varepsilon_n}\lambda\left(\varepsilon_{n+\nu},\varepsilon_{-n-\nu+\mu}\right)
\lambda\left(\varepsilon_n,\varepsilon_{-n+\mu}\right)\nonumber\\
&&\times\left(\varepsilon_n+\varepsilon_{n+\nu}\right)^2
I\left(\varepsilon_n,\Omega_{\mu},\omega_{\nu}\right)\;, \label{MT2}
\end{eqnarray}
and
\begin{eqnarray}
I\left(\varepsilon_n,\Omega_{\mu},\omega_{\nu}\right)&&
=\int\left(dp\right)G_0\left({\bf p},\varepsilon_{n+\nu}\right)
G_0\left({\bf p},\varepsilon_{-n-\nu+\mu}\right)\nonumber\\
&&\times\int\left(dp'\right)G_0\left({\bf p}',\varepsilon_n\right)
G_0\left({\bf p}',\varepsilon_{-n+\mu}\right).\label{MT3}
\end{eqnarray}
Using the same procedure outlined above to calculate the DOS
correction, we get
\begin{eqnarray}
\label{MT4}
\frac{\delta\kappa^{(MT)}}{\kappa_0}=\frac3{\pi^2}\frac{1}{g_T}\frac{g_T\delta}{T_c}\int_{BZ}(dK)\frac{\gamma_{\bf
K}}{\epsilon+z\frac{g_T\delta}{T_c}\left(1-\gamma_{\bf
K}\right)}.
\end{eqnarray}
As expected, the MT correction has the same singular behaviour as
the DOS but opposite sign. On the other hand, because such a
correction involves the coherent tunneling of the fluctuating
Cooper pair from one site to the nearest neighbor, it is
proportional to the lattice structure factor $\gamma_{\bf K}$: due
to this proportionality, in the regime $T-T_c\gg g_T\delta$, the
correction vanishes because $\int_{BZ}(dK)\gamma_{\bf K}\equiv0$.
Let us stress again that this is not the case for the DOS
correction, which in this regime behaves as
$-(1/g)({E_T}/{T_c})(1/\epsilon)$.

\subsection{Aslamazov-Larkin correction}
\begin{figure}
\begin{center}
\includegraphics[width=6cm]{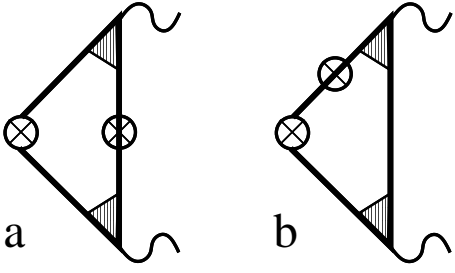}
\end{center}
\caption{Diagrams of the blocks appearing in the Aslamazov-Larkin
contribution to thermal conductivity. (b) has a double
molteplicity, since the bare tunnel vertex can stay on both side
of the block.} \label{figtot1}
\end{figure}
The AL diagrams can be built up by means of blocks in Fig.
\ref{figtot1}, by considering all their possible combinations in
pairs. For a sake of simplicity, we will call the first block,
Fig. \ref{figtot1}(a), $B_1$, and the second one $B_2$. Finally,
one has three different kind of diagrams: the first one, with two
$B_1$-type blocks; the second one with two $B_2$-type blocks, and
the latter, with both of them. Because of the double molteplicity
of $B_2$-type block, one has a total of nine diagrams contributing
to thermal conductivity. In the following, first  we evaluate the
analytical expression of $B_1$ and $B_2$ in the static
approximation, then in the dynamical one, giving the expression of the total AL correction.\\ \indent The general
expression of response function for the AL diagrams reads
\begin{eqnarray}
{\cal
Q}^\mathrm{(AL)}\left(\omega_{\nu}\right)&&=T^2a^2t^4\sum_{\substack{\langle l,i \rangle\\ \langle j,m \rangle} }
\sum_{\Omega_{\mu}}\Lambda_{ij}\left(\Omega_{\mu+\nu}\right)
\Lambda_{ml}\left(\Omega_{\mu}\right)\nonumber\\
&&\times B_\mathrm{left}\left(\omega_{\nu},\Omega_{\mu}\right)
B_\mathrm{right}\left(\omega_{\nu},\Omega_{\mu}\right)\;,
\label{AL1}
\end{eqnarray}
where $B_\mathrm{left}$ and $B_\mathrm{right}$ can be either $B_1$ or
$B_2$-type. \\ \indent $B_1$ block reads
\begin{eqnarray}
B_1\left(\omega_{\nu},\Omega_{\mu}\right)&&=\sum_{\varepsilon_n}\left(
\varepsilon_n+\varepsilon_{n+\nu}\right)
\lambda\left(\varepsilon_{n+\nu},\varepsilon_{\mu-n}\right)
\lambda\left(\varepsilon_n,\varepsilon_{\mu-n}\right)\nonumber\\
&&\times\int(dp)G_0\left( {\bf
p},\varepsilon_{n+\nu}\right)G_0\left( {\bf
p},\varepsilon_{\mu-n}\right)\nonumber\\
&&\times\int(dp')G_0\left( {\bf
p}',\varepsilon_{\mu-n}\right)G_0\left( {\bf
p}',\varepsilon_n\right)\label{AL2}.
\end{eqnarray}
Taking the integrals over the Fermi surface, in the static
approximation, we get
\begin{eqnarray}
B_1\left(\omega_{\nu},0\right)&=&\left(2\pi\nu_F\tau\right)^2
\sum_{\varepsilon_n}\theta\left(\varepsilon_{n+\nu}\varepsilon_n\right)
\left(\varepsilon_n+\varepsilon_{n+\nu}\right)\nonumber\\
&&\times \;\lambda\left(\varepsilon_{n+\nu},-\varepsilon_n\right)
\lambda\left(\varepsilon_n,-\varepsilon_n\right)
\nonumber \\
&=&\left(2\pi\nu_F\right)^2\nonumber \\
&&\hspace{-0.5cm}\times
\left[\sum_{\varepsilon_n<-\omega_{\nu}}+
\sum_{\varepsilon_n>0}\right]
\frac{\varepsilon_n+\varepsilon_{n+\nu}}{\left|
\varepsilon_{n+\nu}+\varepsilon_n\right|}
\frac1{\left|2\varepsilon_n\right|}\label{AL3};
\end{eqnarray}
manipulating the sum, it is easy to see that
\begin{eqnarray}
B_1\left(\omega_{\nu},0\right)&=&\left(2\pi\nu_F\right)^2
\sum_{0<\varepsilon_n<\omega_{\nu}}\frac1{2\varepsilon_n}
\nonumber\\
&=& (2\pi\nu_F)^2\left[\psi\left(\frac{\omega_{\nu}}{2\pi
T}+\frac12\right)-\psi\left(\frac12\right)\right]\nonumber\\
&\approx&\left(\frac{\pi\nu_F}{2}\right)^2\frac{\omega_{\nu}}{T}\label{AL4}\;.
\end{eqnarray}
\indent In the same way as sketched above, one can show, always in
the static approximation, that the block $B_2$ vanishes
identically. Then, all the diagrams containing $B_2$-type blocks
do not give any contribution. Since the only AL diagram with two
$B_1$-type block is proportional to the square of Eq. (\ref{AL4}),
it is quadratic in the external frequency $\omega$, and therefore
vanishes identically in the limit $\omega\rightarrow0$. \\
 \indent To evaluate the first non vanishing AL correction, one has to
consider the dynamical contribution. In such a case, the $B_2$
block, for instance, reads
\begin{eqnarray}
B_2\left(\omega_{\nu},\Omega_{\mu}\right)&&=\sum_{\varepsilon_n}\left(
\varepsilon_n+\varepsilon_{n+\nu}\right)
\lambda\left(\varepsilon_{n+\nu},\varepsilon_{\mu-n}\right)
\lambda\left(\varepsilon_n,\varepsilon_{\mu-n}\right)\nonumber\\
\hspace{-2cm}&&\times\int(dp')G_0\left( {\bf
p}',\varepsilon_{\mu-n}\right)G_0\left( {\bf
p}',\varepsilon_n\right)G_0\left( {\bf
p}',\varepsilon_{n+\nu}\right)\nonumber\\
&&\times\int(dp)G_0\left( {\bf
p},\varepsilon_{n+\nu}\right)\label{AL5}\;.
\end{eqnarray}
\indent In the evaluation of the block, because of the pole
structure of fluctuation propagator, one can neglect the
$\omega_\nu$ dependence\cite{LV04,ISS03}, and keep just the one
in $\Omega_\mu$. The calculation of the integrals and the sums in
the latter equation is, in the dynamical approximation, a little
bit more cumbersome. One has to take into account the different
possible signs of $\Omega_\mu$ and $\varepsilon_n$. Finally, Eq.
(\ref{AL5}) reads
\begin{eqnarray}
\lefteqn{B_2\left(0,\Omega_{\mu}\right)=-2(\pi \nu_F)^2 \sum_{\varepsilon_n}
\frac{2\varepsilon_n}{(2\varepsilon_n-\Omega_\mu)^2}
[\theta(\Omega_\mu)(\theta(\varepsilon_n-\Omega_\mu)}\hspace{0.7cm}\nonumber \\
&&+\theta(-\varepsilon_n))+
\theta(-\Omega_\mu)(\theta(\Omega_\mu-\varepsilon_n)+\theta(\varepsilon_n))]\;,\label{AL6}
\end{eqnarray}
$\theta(x)$ being the step function.\\
\indent By taking the lowest order in the bosonic frequency
$\Omega_\mu$, one gets the result for the block
\begin{equation}
B_2\left(0,\Omega_{\mu}\right)=-\frac{1}{2}\left(\frac{\pi\nu_F}{2T}\right)^2\Omega_\mu\;.\label{AL7}
\end{equation}
\indent In the same way, one can evaluate also $B_1$ with the result
\begin{equation}
B_1\left(0,\Omega_{\mu}\right)=-2B_2\left(0,\Omega_{\mu}\right)\;,\label{AL8}
\end{equation}
which is consistent with the homogeneous case\cite{LV04,ISS03}. The sum
over $\Omega_\mu$ in the response function can be performed by
writing the sum as an integral\cite{LV04}, and exploiting the
properties of  the pair correlators.\\
\indent Finally, the AL dynamical correction to thermal
conductivity reads
\begin{equation}
\frac{\delta \kappa^{(AL)}}{\kappa_0}=\frac{9}{2\pi}
\frac{1}{g_T}\left( \frac{g_T\delta}{T_c}\right)^2\int_{BZ}(dK)
\frac{(1-\gamma_{\bf k})^2}{\epsilon+z\frac{g_T\delta}{T_c}
\left(1-\gamma_{\bf K}\right)}\;.\label{AL9}
\end{equation}
\indent The latter equation is the first non vanishing correction due
to AL channel. Such a correction is always positive, and it
depends, as in the MT, on the lattice structure factor
$\gamma_{\bf K}$, but it does not vanishes in the regime $T-T_c\gg
g_T \delta$. This is a good feature of the system, since far from
$T_c$, the dynamical contribution plays an important role, and in
this region, one has to compare it with DOS one, as discussed in
the following section. Here, we just observe that since the
corrections, Eqs. (\ref{23}), (\ref{MT4}) and (\ref{AL9}), have
different signs, nonmonotonic behaviour in the total correction
is expected, depending on the ratio ${g_T\delta}/{T_c}$.

\section{Discussion}

\begin{figure}[tp]
\begin{center}
\includegraphics[width=9cm]{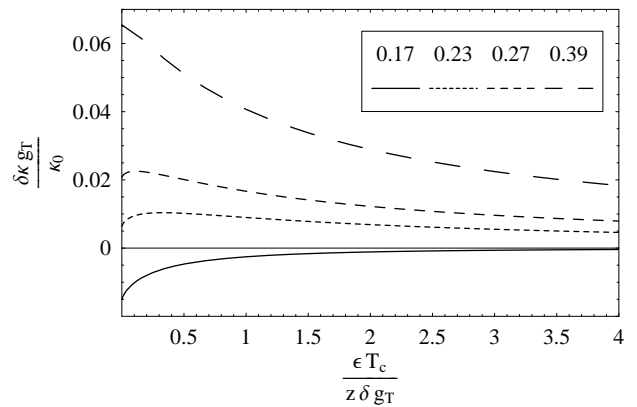}
\end{center}
\caption{Total fluctuation correction to the thermal conductivity
for different values of ${g_T\delta}/{T_c}$ for a
two-dimensional system. A $1/\epsilon$ suppression is observed
at high temperatures, with a sign depending on such ratio. At low
temperatures, a finite correction, inversely proportional to the
coordination number $z$, is reached at $\epsilon=0$. In a finite
interval of values of $g_T\delta/T_c$, a nonmonotonic behaviour
of the correction is observed, where the correction is positive
and increasing with decreasing temperature, reaches a maximum and
then goes to a smaller (possibly negative) value at the critical
temperature.} \label{fig7}
\end{figure}

As we have seen, the total superconducting fluctuation correction
to the thermal conductivity close to critical temperature is given
by the following expression
\begin{equation}
\label{TOT1}
\frac{\delta\kappa}{\kappa_0}=\frac3{\pi^2}\frac{\delta}{T_c}
\int_{BZ}(dK)\frac{\left(1-\gamma_{\bf
K}\right)\left[\frac{3\pi}2\frac{g_T\delta}{T_c}\left(1-\gamma_{\bf
K}\right)-1\right]}{\epsilon+z\frac{g_T\delta}{T_c}\left(1-\gamma_{\bf
K}\right)}.
\end{equation}
This correction has been obtained at all orders in the tunneling
amplitude in the ladder approximation. Its behaviour is plotted in
Fig. \ref{fig7} as a function of the reduced temperature for
the case of a two dimensional sample, and for different
values of the ratio $g_T\delta/T_c$. We can recognize two
different regimes of temperatures: far from $T_c$, $\epsilon\gg
{g_T\delta}/{T_c}$, and close to $T_c$, $\epsilon\ll
{g_T\delta}/{T_c}$. For a sake of simplicity, we will identify
these two regimes as ``high temperatures'' and ``low
temperatures'', respectively.

\begin{itemize}

\item {\bf High temperature regime} $\epsilon\gg
{g_T\delta}/{T_c}$. In this region, the electrons do not
tunnel efficiently between the grains and the system behaves 
almost as an ensemble of zero-dimensional systems. As a
consequence, only the DOS and AL terms contribute
significantly to the superconducting fluctuations; the correction
to heat conductivity reads
\begin{eqnarray}
\label{TOT2}
\frac{\delta\kappa}{\kappa_0}&\approx&\frac3{\pi^2}\frac{\delta}{T_c}
\frac1\epsilon
\left[\frac{3\pi}2\frac{g_T\delta}{T_c}\left(1+\frac1{z}
\right)-1\right]\;.
\end{eqnarray}
 This expression shows a $1/\epsilon$ singularity and it can
have either positive or negative sign, depending on the ratio
$g_T\delta/T_c$; we call $\gamma_1$ the value of the above-mentioned ratio solution of Eq. (\ref{TOT2}). 
 In the absence of renormalization due to tunnelling, the correction
is negative and corresponds to the typical singularity of the
quasi-zero-dimensional density of state. On the other hand,
increasing the barrier transparency $g_T\delta$, the correction
grows due to the presence of the direct channel, i.e., the AL
term, which becomes more and more important, until the correction
itself vanishes at $\gamma_1$, after which it becomes positive. A
direct comparison with the behaviour of the electrical
conductivity\cite{BEL00} shows that, already at this level, there
is a positive violation of the Wiedemann-Franz law, being
\begin{eqnarray}
\frac{\delta L}{L_0}&=&\frac{\delta\kappa}{\kappa_0}-
\frac{\delta\sigma}{\sigma_0}\nonumber\\
&\approx&\left[-\frac3{\pi^2}+\frac9{2\pi}\frac{g_T\delta}{T_c}
\frac{z+1}z +\frac{7\zeta(3)}{\pi^2}\right]
\frac\delta{T_c}\frac1\epsilon\label{WF1}
\end{eqnarray}

\item{\bf Low temperature regime} $\epsilon\ll
{g_T\delta}/{T_c}$. Here the tunneling is effective and there
is a crossover to the typical behaviour of a homogeneous system,
as $T\rightarrow T_c$, from the point of view of the fluctuating
Cooper pairs. Physically, the bulk behaviour is recovered, and one
gets a non divergent (though nonanalytic) correction even at
$\epsilon=0$, where it equals
\begin{equation}
\label{TOT3}
\frac{\delta\kappa\left(\epsilon=0\right)}{\kappa_0}=\frac3{z\pi^2}\frac1{g_T}
\left(\frac{3\pi}2\frac{g_T\delta}{T_c}-1\right)\;.
\end{equation}
The latter equation gives the saturation value in any dimension;
it is also evident the $1/{g_T}$ order of the perturbation
theory. Again, the value of the constant can be either
negative or positive. The correction vanishes at a value $g_T\delta/T_c=\gamma_2$ 
which is independent on the dimensionality and larger than
$\gamma_1$. In the interval
$\gamma_1<g_T\delta/T_c<\gamma_2$, it has a non-monotonic behaviour,
being positive and increasing for high temperatures and negative
for low temperatures. Such a behaviour has been represented, for the
case of $d=2$, in Fig.\ref{fig7}.
The deviation from the Wiedemann-Franz law in the low temperature
region is much more evident than in the high temperature one,
because of the pronounced singular behaviour of the electrical
conductivity close to the critical temperature\cite{BEL00}.

\end{itemize}

\section{Conclusions}

We have calculated the superconducting fluctuation
corrections to heat conductivity. In the region of temperatures
$T-T_c\gg g_T\delta$, a strong singular correction is found,
reported in Eq. (\ref{TOT2}), corresponding to the sum of the DOS
renormalization and the AL contribution in a
quasi-zero-dimensional system. Moving closer to the critical
temperature, when $T-T_c\ll g_T\delta$, the divergent behaviour of
the DOS term is cut off by the MT correction, which has opposite
sign, while the AL term regularizes by itself to a finite value;
this regularization signals the fact that the system undergoes a
crossover to the homogeneous limit. A nondivergent behaviour is
found at the critical temperature, in agreement with previous
calculation in homogeneous superconductors\cite{NS02,USH02}. The
energy scale that separates the two regions, $g_T\delta$, can be
recognized as the inverse tunneling time for a single
electron\cite{BEAH01}. As a final remark, we want to note
that the ratio $zg_T\delta/T_c$ appears as the coefficient of the
${\bf K}-$dependent term in the superconducting fluctuation
propagator, Eq. (\ref{18}): from the standard theory of the
superconducting fluctuations, the coefficient of $K^2$ in the
propagator is actually the superconducting coherence
length\cite{LV04}; we can therefore define an ''effective
tunneling superconducting coherence length'' as
$\xi_0^{(T)}=a\sqrt{g_T\delta/T_c}$. From this definition, we can
see that, if $\xi_0^{(T)}\ll a$, the grains are strictly
zero-dimensional at high-temperature and the correction to the
thermal conductivity is always negative, while if $\xi_0^{(T)}\gg
a$, the direct channel of the superconducting correlations is
strong enough to change sign to such correction.

We gratefully acknowledge illuminating discussions with A. A.
Varlamov, I.V. Lerner and I.V. Yurkevich. This work was supported
by IUF (FWJH), and Universit\'e franco-italienne (R. Ferone).

\appendix

\section{Microscopic derivation of fluctuation propagator \label{A}}

Here we report a short description of the derivation of Eq.(\ref{18}),
evaluated in Ref. \cite{BEL00}, to remind the reader the main steps and the main
assumptions of the calculation. We start from the expression of
the partition function in the interaction representation
\begin{eqnarray}
{\cal Z}&=&{\rm Tr}\;{\rm exp}\left({-\int_0^{\beta}\hat
H(\tau)d\tau}\right)
\nonumber\\
&&={\rm Tr}\;\Bigg\{{\rm exp}\left({-\int_0^{\beta}\hat
H_0(\tau)d\tau}\right)\nonumber\\
&&\times T_{\tau}\; {\rm exp}\left({-\int_0^{\beta}\left[\hat
H_P(\tau)+\hat H_T\right]d\tau}\right)\Bigg\}\;.\label{13}
\end{eqnarray}
We decouple the electronic fields in $\hat H_P$ by means of
Hubbard-Stratonovich transformation, introducing the order
parameter field $\Delta$; because of our assumption,
$E_T\gg\Delta$, the grains can be considered strictly
zero dimensional and we can neglect the spatial coordinate
dependence in the field $\Delta_i$ in Eq. (\ref{13}). We now
expand over the field $\Delta_i$; the expansion is justified by
our assumption to be close but above to the critical temperature
where the mean field (BCS) value of order parameter is still zero;
moreover, we have to expand the action to the second order in $t$,
too; this expansion is justified in the region~\cite{LY00}
$t\ll 1/\tau\ll E_T$. We obtain two different contributions to
the action: the first one is the typical action of superconducting
fluctuations; the other one is the tunneling correction:
$S_\mathrm{eff}=S_\mathrm{eff}^0+S_\mathrm{eff}^t$. The first term is\cite{LV04}
\begin{eqnarray}
\lefteqn{S_\mathrm{eff}^0=-\frac
TV\sum_{\Omega_{\mu}}\left|\Delta_i\left(\Omega_{\mu}\right)\right|^2}
\nonumber\\
&&\times\Bigg[\frac1\lambda-4\pi\nu_FT{\rm
\tau}\sum_{2\varepsilon_n>\Omega_{\mu}}\lambda\left(\varepsilon_n,\varepsilon_{\mu-n}
\right)\Bigg]\;. \label{14}
\end{eqnarray}
$\Omega_{\mu}$ always appears as the combination of two fermionic
Matsubara frequencies and it is therefore a bosonic one, as it
should be. The sum over the fermionic frequencies in Eq.
(\ref{14}) is logarithmically divergent and must be cut off at
Debye's frequency\cite{LV04}; using the definition of
superconducting critical temperature, one obtains
\begin{eqnarray}
\lefteqn{S_\mathrm{eff}^0=-\nu_F\frac TV
\sum_{\Omega_{\mu}}\left|\Delta_i\left(\Omega_{\mu}\right)\right|^2}
\nonumber\\
&&\times\Bigg[\ln\frac
T{T_c}+\psi\left(\frac12+\frac{\left|\Omega_{\mu}\right|}{4\pi
T_c}\right) -\psi\left(\frac12\right)\Bigg] \;,\nonumber
\end{eqnarray}
where $\psi(x)$ is the digamma function, defined as the
logarithmic derivative of gamma function\cite{GR94,LV04}. Close to
critical temperature, $T\simeq T_c$, as already mentioned, the
main contribution to singular behaviour comes from ''classical"
frequencies, $\left|\Omega_{\mu}\right|\ll T_c$. Then, we can
expand the $\psi$ function in the small parameter
${\left|\Omega_{\mu}\right|}/{T_c}$:
\begin{eqnarray}
S_\mathrm{eff}^0=-\nu_F\frac TV \sum_{{\bf K},\Omega_{\mu}}\left[\ln\frac
T{T_c}+\frac{\pi\left|\Omega_{\mu}\right|}{8T_c}\right]
\left|\Delta_{\bf K}\left(\Omega_{\mu}\right)\right|^2\;.
\label{15}
\end{eqnarray}
\indent In the last expression, for later convenience, we
considered the lattice Fourier transform: ${\bf K}$ belongs to the
first Brillouin zone of reciprocal grain lattice. As it has
been mentioned, the zero-dimensional character of the grain
resides in the independence of the action on coordinates
inside each grain.

\begin{figure}[]
\begin{center}
\includegraphics[width=8cm]{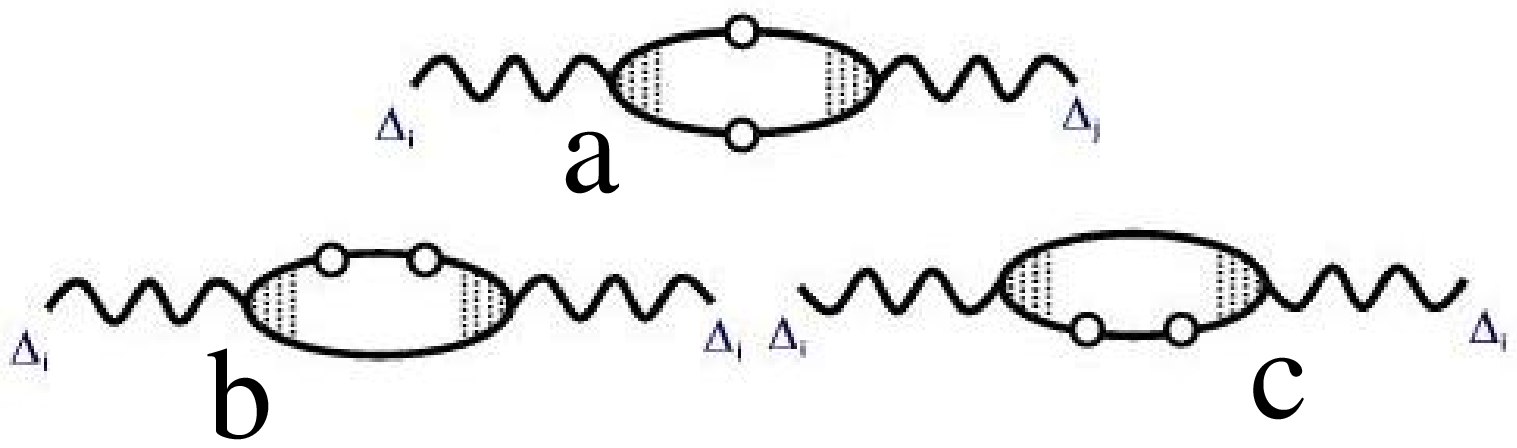}
\includegraphics[width=8cm]{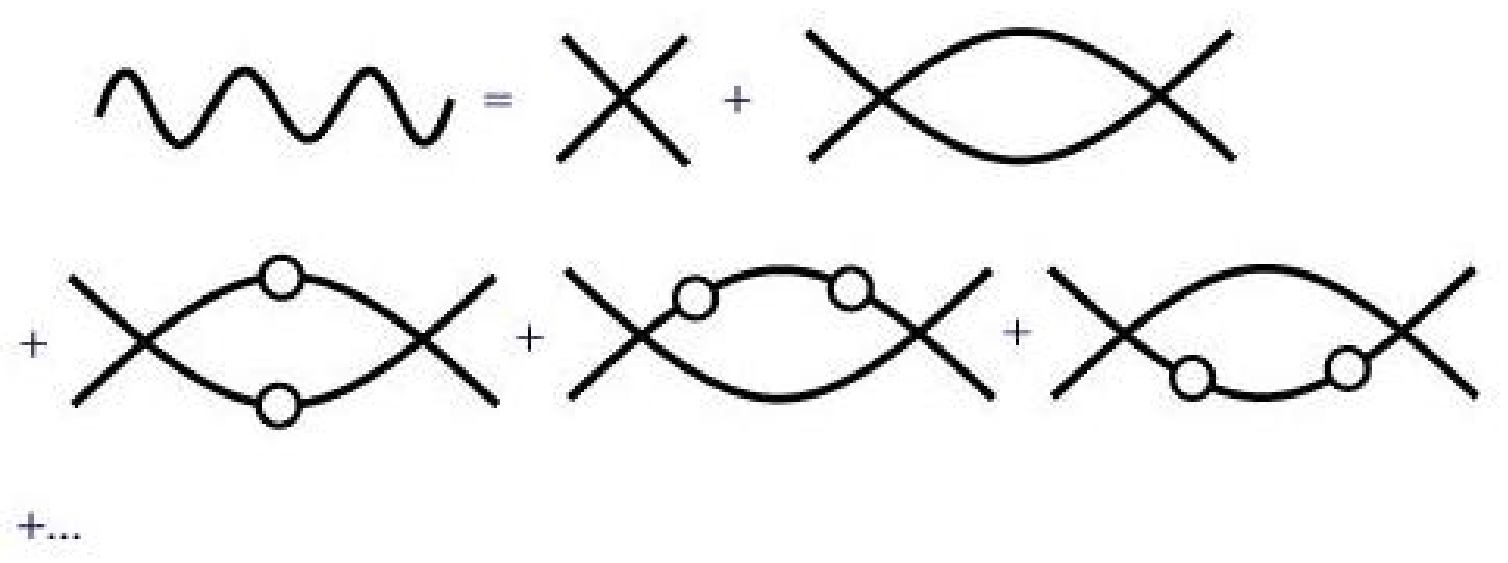}
\end{center}
\caption{Top: The total tunneling correction to the fluctuation
propagator self-energy is reported. The upper diagram is related
to the possibility of tunneling of both electrons forming the
fluctuating Cooper pair during the lifetime $\tau_{GL}$ of the
Cooper pair itself. The other two diagrams consider the
renormalization of the intragrain fluctuation propagator. Bottom:
The ladder series for the fluctuation propagator in the presence
of tunneling is reported. The crosses are BCS electron-electron
interaction.} \label{fig2}
\end{figure}

The tunneling-dependent part of the action is calculated starting
from diagrams in Fig.\ref{fig2}: they represent the first
non-vanishing correction to fluctuation propagator due to
tunneling. Their reexponentiation corresponds to the sum of the ladder
series of tunneling and pairing interaction as reported in
Fig.\ref{fig2}. The calculation of diagram $a$, Fig.\ref{fig2}(a), gives the contribution due to the possibility of tunneling
of both electrons during the lifetime of the fluctuating Cooper
pair, i.e. the Ginzburg-Landau time
$\tau_{GL}=\pi/{8\left(T-T_c\right)}$; it is equal to
\begin{eqnarray}
S_\mathrm{eff}^{t,(a)}=zg_T\sum_{{\bf
K},\Omega_{\mu}}\gamma_{\bf K}\left|\Delta_{\bf
K}\left(\Omega_{\mu}\right)\right|^2\;,
\label{16}
\end{eqnarray}
where, as mentioned, $z$ is the number of nearest neighbors.\\
\indent Figures \ref{fig2}(b) and \ref{fig2}(c),
give an identical contribution, which is related to the
probability that a single electron, participating in the
fluctuating Cooper pair, undergoes a double tunneling, back and
forth, during the Ginzburg-Landau time. Such a contribution reads
\begin{eqnarray}
\label{17} S_\mathrm{eff}^{t,(b+c)}=-zg_T\sum_{{\bf
K},\Omega_{\mu}}\left|\Delta_{\bf
K}\left(\Omega_{\mu}\right)\right|^2.
\end{eqnarray}
The final result for fluctuation propagator at every order in
tunneling in the ladder approximation is
\begin{eqnarray}
\label{A18} \Lambda_{\bf
K}\left(\Omega_{\mu}\right)=-\frac1{\nu_F} \frac1{\ln\frac
T{T_c}+\frac{\pi\left|\Omega_{\mu}\right|}{8T_c}+z\frac{g_T\delta}{T_c}\left(1-\gamma_{\bf K}\right)}.
\end{eqnarray}

Finally, we notice that the classical limit ($\Omega_{\mu}=0$) for the fluctuation
propagator Eq.(\ref{A18}) can be obtained from a straightforward generalization
of the Ginzburg-Landau functional for granular metals
\begin{eqnarray}
&&{\cal F}\left[\Psi\right]=\sum_{\left\langle i,j\right\rangle}
\int_{i}(d{\bf r})\int_{j}(d{\bf
r}^\prime)\Big[a\Psi_i^*\left({\bf r}\right)\Psi_j\left({\bf
r}^\prime\right)\delta_{ij}\delta\left({\bf r}-{\bf
r}^\prime\right)\nonumber\\
&&+{\cal J}\left|\Psi_i\left({\bf r}\right)-\Psi_j\left({\bf
r}^\prime\right)\right|^2\Big]\label{LD1}
\end{eqnarray}
where the parameter $a$ is given by $(1/{4m\xi^2})\ln(T/{T_c})$, where $m$ is the electron mass, while the so-called
Josephson parameter keep track of the tunneling effect: ${\cal
J}=(1/{4m\xi^2})({zg_T\delta}/{T_c})$. See also
Ref. \onlinecite{DIG74} for the region of applicability of the
theory reported above.


\begin{references}
\bibitem{nota3} We assume that the tunneling
contacts are {\em not} point-like. If they were (\onlinecite{LY04,CML96}),
the tunneling conductance of the grains would be renormalized to
$g_T^\mathrm{(ren)}\propto{g_T}/{\left(1+g_T\right)^2}$ by the
backscattering: this fact is immediately understood saying that in
this situation the average area of a tunneling contact is
$A^2\approx\lambda_F^2$ where $\lambda_F$ is the Fermi wavelength
and no more than one channel can be available for the transport.
One is consequently forced to take into account higher-order
diagrams in tunneling even in the Green's function. On the other
hand, if $\lambda_F\ll A\ll a$, higher-order diagrams will be
smaller by a factor ${g_T}/g$ and in our approximation they
can be safely neglected.
\bibitem{nota1} The tunneling
renormalization of the cooperon vertex correction is mandatory in
the calculation of other thermodynamic or transport quantities,
such as the correction to the superconducting critical
temperature\cite{BELV04} or the superconducting fluctuation
corrections to the electrical conductivity in the absence of the
magnetic field\cite{BEL00}, in order to avoid spurious
divergences. In the case of the heat conductivity, such a
renormalization gives only a small correction of the order of
${\cal O}\left(\tau g_T\delta\right)$ and can be safely neglected.
\end{references}
\end{document}